\documentclass[twocolumn,aps,preprintnumbers,amsmath,amssymb,superscriptaddress,nofootinbib]{revtex4}

\usepackage{graphicx,subfigure,multirow}
 \usepackage{longtable}
\usepackage{dcolumn}
\usepackage{bm}
\usepackage[all]{xy}
\usepackage{color}
\usepackage[usenames,dvipsnames]{xcolor}
\usepackage[unicode=true,pdfusetitle,
 bookmarks=true,bookmarksnumbered=false,bookmarksopen=false,citecolor=Turquoise,
 breaklinks=false,pdfborder={0 0 1},backref=false,colorlinks=true,pdfpagemode=FullScreen]
 {hyperref}

\makeatletter

\newcommand{\fmslash}[2][0mu]{%
  \mathchoice
    {\fmsl@sh\displaystyle{#1}{#2}}%
    {\fmsl@sh\textstyle{#1}{#2}}%
    {\fmsl@sh\scriptstyle{#1}{#2}}%
    {\fmsl@sh\scriptscriptstyle{#1}{#2}}}
\newcommand{\fmsl@sh}[3]{%
  \m@th\ooalign{$\hfil#1\mkern#2/\hfil$\crcr$#1#3$}}
\makeatother

\newcommand{\lsim}{{\;\raise0.3ex\hbox{$<$\kern-0.75em\raise-1.1ex\hbox{$\sim$}}\;}}
\newcommand{\gsim}{{\;\raise0.3ex\hbox{$>$\kern-0.75em\raise-1.1ex\hbox{$\sim$}}\;}}

\newcommand{\beq}{\begin{equation}}
\newcommand{\eeq}{\end{equation}}
\newcommand{\bea}{\begin{eqnarray}}
\newcommand{\eea}{\end{eqnarray}}
\mathchardef\minus="002D

\newcommand{\bb}{\textrm{b}_b}
\newcommand{\bc}{\textrm{b}_c}
\newcommand{\bs}{\textrm{b}_s}
\newcommand{\cb}{\textrm{c}_b}
\newcommand{\cc}{\textrm{c}_c}
\newcommand{\cs}{\textrm{c}_s}
\newcommand{\col}{\textcolor{black}}
\begin{document}

\preprint{
\begin{minipage}[b]{1\linewidth}
\begin{flushright}
APCTP-PRE2015-016\\
 IPMU15-0088 \\
 CETUP2015-012
 \end{flushright}
\end{minipage}
}

\title{Enhancement of new physics signal sensitivity with mistagged charm quarks}

\author{Doojin Kim} 
\email[]{immworry@ufl.edu}
\affiliation{Physics Department, University of Florida, Gainesville, FL 32611, USA}
\author{Myeonghun Park}
\email[]{parc@apctp.org}
\affiliation{ Asia Pacific Center for Theoretical Physics, San 31, Hyoja-dong, \\
  Nam-gu, Pohang 790-784, Republic of Korea}
  \affiliation{Department of Physics, Postech, Pohang 790-784, Korea}
\affiliation{Kavli IPMU (WPI), The University of Tokyo, Kashiwa, Chiba 277-8583, Japan}

\date{\today}

\begin{abstract}
We investigate the potential for enhancing search sensitivity for signals having charm quarks in the final state, using the {\it sizable} bottom-mistagging rate for charm quarks at the LHC. Provided that the relevant background processes contain light quarks instead of charm quarks, the application of $b$-tagging on charm quark-initiated jets enables us to reject more background events than signal ones due to the relatively small mistagging rate for light quarks. The basic idea is tested with two rare top decay processes: i) $t\rightarrow c h\rightarrow c b\bar{b}$ and ii) $t\rightarrow b H^+ \rightarrow b\bar{b}c$ where $h$ and $H^+$ denote the Standard Model-like higgs boson and a charged higgs boson, respectively. The major background source is a hadronic top quark decay such as $t\rightarrow bW^+ \rightarrow b \bar{s}c$. We test our method with Monte Carlo simulation at the LHC 14TeV, and find that the signal-over-background ratio can be increased by a factor of $\mathcal{O}(6-7)$ with a suitably designed (heavy) flavor tagging algorithm and scheme. 

\end{abstract}


\maketitle

\section{Introduction} 
The discovery of the Higgs particle at the Large Hadron Collider (LHC)~\cite{Aad:2012tfa,Chatrchyan:2012ufa} reaffirms that the Standard Model (SM) is a successful description of fundamental particles and their interactions in nature. Nevertheless, the detailed mechanism of protecting its mass scale from large quantum corrections is still unexplained by the SM, and new physics beyond the Standard Model (BSM) is anticipated to address this puzzle. Since the corrections are dominantly contributed by the top quark, the top quark sector has been regarded as a promising host to accommodate and reveal new physics signatures. Furthermore, the LHC, dubbed a ``top factory'', is capable of copiously producing top quarks in pairs via the strong interaction, and it can therefore be taken as a great venue to discover new physics phenomena using top quarks.

We emphasize that although many physical properties of the top quark have been measured with great precision since its discovery, its decays are relatively poorly-measured; typical errors in the top quark decays are of $\mathcal{O}(10\%)$ mostly coming from systematics in the measurement of the $t$-channel single top quark cross section~\cite{Agashe:2014kda,Khachatryan:2014nda}. Hence, any new physics effects emerging in the top quark decay channels are, in principle, less constrained by current experimental data. 

One of the rare top decay examples to be considered here is $t \rightarrow ch$ via a flavor changing neutral current (FCNC)~\cite{Eilam:1990zc,Mele:1998ag,AguilarSaavedra:2004wm,Craig:2012vj} followed by the dominant decay mode of the higgs of 125 GeV, i.e., $h \rightarrow b\bar{b}$. In principle, nothing precludes the top quark from decaying in this manner. Nevertheless, the SM prediction on the branching ratio (Br) of this process is extremely small due to the famous Glashow-Iliopoulos-Maiani mechanism and second-third generation mixing suppression, which results in Br$(t\rightarrow ch)_{\textnormal{SM}}\approx 10^{-13}-10^{-15}$~\cite{Eilam:1990zc,Mele:1998ag,AguilarSaavedra:2004wm}. Therefore, a significant excess from such a small SM expectation could be a convincing sign of the existence of new physics. In fact, once new physics is introduced, the aforementioned suppressions can be relaxed, and thus fairly larger branching fractions can be anticipated, e.g., Br$(t\rightarrow ch)_{\textnormal{BSM}}\approx 10^{-3}-10^{-6}$ depending on the details of the BSM models of interest~\cite{AguilarSaavedra:2004wm}, which is comparable with the recent experimental bound reported by the CMS collaboration~\cite{CMStch}.	

Another exciting scenario to be considered here is $t \rightarrow b H^+$ where the charged higgs sequentially decays into a charm quark and an anti-bottom quark unlike the typical decay mode of $H^+ \rightarrow c\bar{s}$. A sizable branching fraction of $H^+\rightarrow c \bar{b}$ arises in a few models with two or more higgs doublets: for example, multi-higgs doublet models (MHDM)~\cite{Grossman:1994jb}, flipped two-higgs doublet models (2HDM)~\cite{Akeroyd:1994ga,Logan:2010ag,Aoki:2009ha} with ``natural flavor conservation'', and Aligned-2HDM~\cite{Pich:2009sp}. Depending on the model details, Br$(H^+\rightarrow c\bar{b})$ could be as large as $\sim 80\%$~\cite{Akeroyd:2012yg}. Although existing experimental searches of $t\rightarrow bH^+ \rightarrow b\bar{s}c$ done by the CDF~\cite{Aaltonen:2009ke} and ATLAS~\cite{Aad:2013hla} collaborations could be applied to the decay of $H^+\rightarrow c\bar{b}$~\cite{Logan:2010ag}, an enhanced branching ratio motivates more dedicated searches to discover a new phenomenon or directly constrain the parameter space in the relevant physics models. 

For the purpose of concreteness we focus on the collider signatures in the context of pair-produced top quarks, and assume that one of the top quarks decays into two bottom and one charm quarks via the decay sequences described above while the other follows the regular leptonic decay cascade. Provided with the visible final state defined by the signal processes, obviously, the dominant SM background is semi-leptonic top quark pair production. Since there exist three bottom quarks for the signal process vs. two bottom quarks for the background one, the requirement of three bottom-tagged jets can substantially reduce background events. It is noteworthy that this event selection enables us to have the hadronic top quark decaying into $b\bar{s}c$ (i.e., $t\rightarrow bW^+ \rightarrow b\bar{s}c$) as a main background source because the $b$-mistagging rate for charm quarks is rather sizable. We henceforth take it as the major background unless specified otherwise.

We point out that, remarkably enough, the high mis-tagging rate for charm quarks can be useful for a further improvement in the relevant signal-over-background ratio ($S/B$). More specifically, if one demands an additional bottom-tagged jet, then signal events can be selected by tagging the remaining charm quark as a bottom quark, whereas background events can be selected by tagging the remaining strange quark as a bottom quark for which the corresponding mis-tagging rate is typically far smaller than that for charm quarks. Therefore, we expect that the relevant signal sensitivity gets increased so that it is possible to probe smaller branching fractions of signal processes.\footnote{In general, the basic idea can be applied to the cases where the signal comes along with charm-induced jet(s) in the final state, whereas the counterparts in the background are light quark-induced jet(s).} Of course, a non-negligible reduction in the signal acceptance due to the additional $b$-jet requirement could be an issue. Given an immense production cross section of top pairs and a large expected integrated luminosity, for example, $\mathcal{L}=300\textrm{fb}^{-1}$ at the 14TeV LHC, adequate statistics can be nevertheless achieved in these search channels.     

\section{Expected enhancement and potential issues}
To develop intuition on the basic idea described thus far, we provide a rough estimation of the expected enhancement by parametrizing pertinent efficiencies. As mentioned before, a way to enhance the $S/B$ (before any posterior analysis using kinematic variables) is to require one more $b$-tagged jet in the final state, utilizing the sizable mistagging rate of charm-induced jets. For more systematic comparison, we begin with (would-be) conventional selection scheme (denoted by 3b), that is, three bottom jets, one regular jet, and a $W$ gauge boson. Since the $W$ is irrelevant to the later discussion, we drop it for convenience. We first define some of the efficiencies with respect to the identification of bottom-initiated jets; $\bb$ as $b$-tagging efficiency of $b$ quark, $\bc$ as $b$-mistagging efficiency of $c$ quark, and $\bs$ as $b$-mistagging efficiency of $s$ quark (light quarks). With this set of definitions and $S$ being the number of signal events before the tagging procedure, the expected number of signal events in the 3b scheme ($S_{\textnormal{3b}}$) is given by 
\bea
S_{\textnormal{3b}} = S \left\{\bb^3(1-\bc)+3\bb^2\bc(1-\bb)\right\} \label{eq:S3b}\, ,
\eea
where the first term represents the leading contribution while the second term represents the subleading contribution such as the case where $c$-induced jet is mistagged while one of the $b$-induced jets is not tagged. When it comes to the major background, the leading contribution comes from a hadronic $W$ decaying into $c$ and $s$ as mentioned earlier. Therefore, the expected number of background events in the 3b scheme ($B_{\textnormal{3b}}$) is
\bea
B_{\textnormal{3b}} = B \bb^2\left\{\bc(1-\bs)+\bs(1-\bc)+\frac{2\bc\bs}{\bb}(1-\bb) \right\} ,
 \label{eq:B3b}
\eea
where $B$ denotes the number of background events before the tagging procedure. We then have the $S/B$ in the 3b scheme as
\bea
\left(\frac{S}{B}\right)_{\textnormal{3b}}=\left(\frac{S}{B} \right) \frac{\bb(\bb-4\bb\bc+3\bc)}{\bb\bc+\bb\bs+2\bc\bs-4\bb\bc\bs}.
\eea
Here we assume that the expected number of background events originating from $t\rightarrow b\bar{d}u$ is negligible because two light quarks are involved. 

On the other hand, if we modify the aforementioned selection scheme by requiring an additional $b$-tagged jet instead of a regular jet (denoted by 4b), the expected numbers of signal and background events ($S_{\textnormal{4b}}$ and $B_{\textnormal{4b}}$, respectively) are expressed as
\bea
S_{\textnormal{4b}} &=& S \bb^3\bc\, ,
 \label{eq:sms} \\
B_{\textnormal{4b}} &=& B \bb^2\bc\bs \, ,
\eea
from which the relevant $S/B$ is simply given by
\bea
\left(\frac{S}{B}\right)_{\textnormal{4b}}=\left(\frac{S}{B}\right) \frac{\bb}{\bs}\, ,
\eea
where a small portion from $t\rightarrow b\bar{d}u$ is neglected again in computing $B_{\textnormal{4b}}$. Defining the improvement of the 4b scheme with respect to the 3b scheme as $I$, we have
\bea
I=\frac{(S/B)_{\textnormal{4b}}}{(S/B)_{\textnormal{3b}}}=\frac{\bb\bc+\bb\bs+2\bc\bs-4\bb\bc\bs}{\bs(\bb-4\bb\bc+3\bc)}.
\eea

\begin{figure}[t]
\centering
\includegraphics[width=0.41\textwidth]{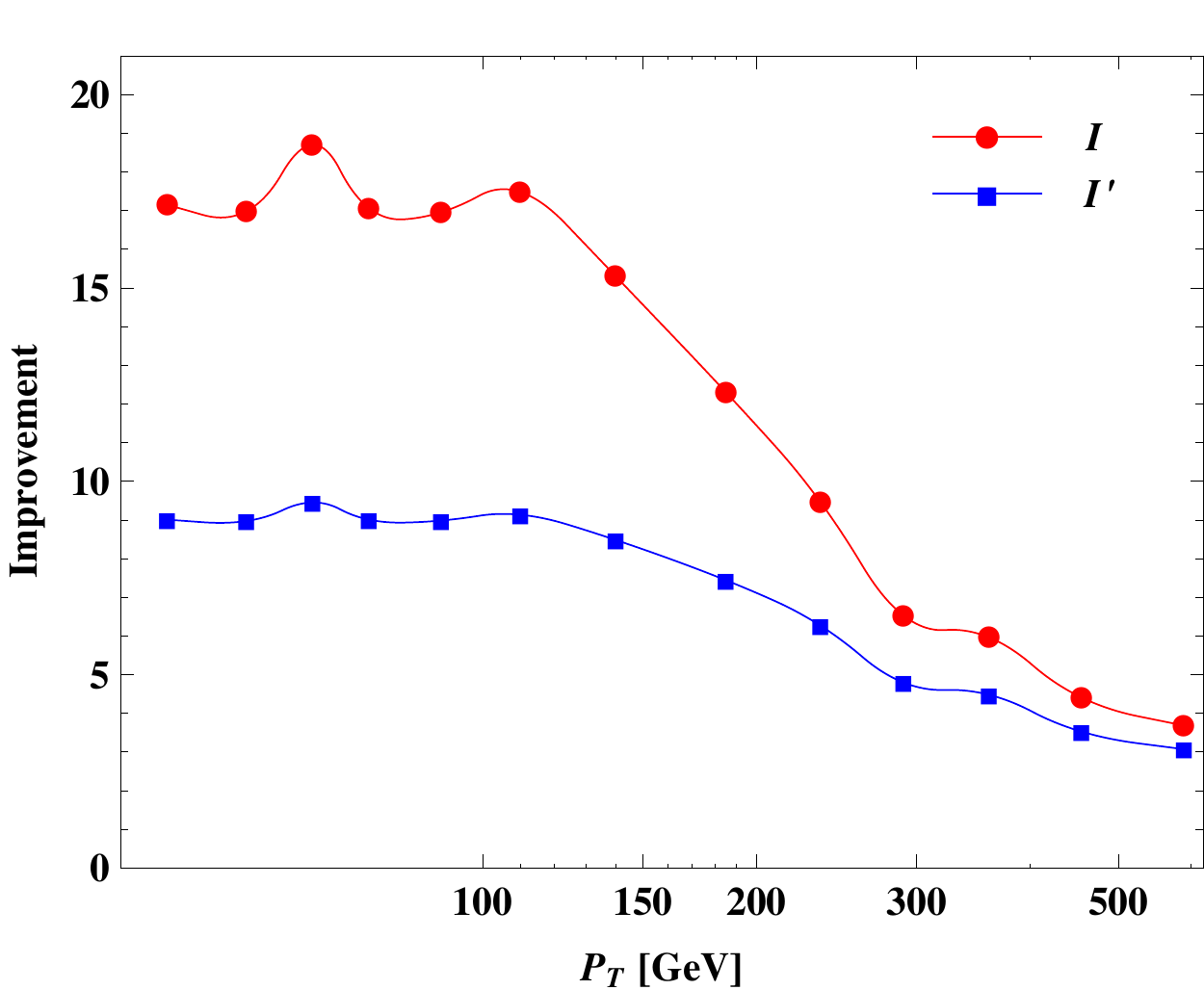}
\caption{\label{fig:improvements} Potential improvements $I$ and $I'$ of signal-over-background ratio as a function of $P_T$ of jets in $t\bar t$ events. In this plot, the dependence on the rapidity of a jet is integrated out ($|\eta|<2.4$). The efficiencies associated with $b$-tagging are obtained from Ref.~\cite{Chatrchyan:2012jua}, whereas for $I'$ the efficiencies associated with $c$-tagging are fixed to be the average values of the medium operating point in Ref.~\cite{ATLASctag} for which $(\cb, \cc, \cs)\approx(0.125,0.20,0.010)$.}
\end{figure}

Since tagging efficiencies vary in the transverse momentum of jets, it is interesting to investigate the dependence of $I$ according to the $P_T$ of $b$-jets, which is explicitly shown by red dots in FIG.~\ref{fig:improvements}. As an example tagging scheme, the CSVM tagger of the CMS collaboration has been adopted, and relevant efficiencies are applied based upon the values reported in Ref.~\cite{Chatrchyan:2012jua} wherein they have measured the data using $t\bar{t}$ events. To understand this behavior more intuitively, it is worthwhile to rewrite $I$ using the leading contributions (i.e., the first terms) in eqs.~(\ref{eq:S3b}) and~(\ref{eq:B3b}):
\bea
I\approx \frac{\bc(1-\bs)}{\bs(1-\bc)}.\label{eq:Iapp}
\eea
One can easily see that the value of $I$ is solely governed by $\bc$ and $\bs$. More specifically, this is an increasing function as $\bc$ ($\bs$) increases (decreases) so that a large gap between $\bc$ and $\bs$ is favored to attain a large improvement. In fact, it turns out that heavy flavor quarks are less tagged as $b$-jets while light quarks fake $b$-jets more often in the high $P_T$ region. The reason is that jets with a large $P_T$ are typically collimated so that errors in particle tracking, which is involved in the tagging algorithm, are likely to increase. As a consequence, less significant improvement is shown in the high $P_T$ region. 

A couple of issues may arise in this strategy. Now that charm quark tagging techniques are being developed~\cite{ATLASctag}, one could apply it to improve the $S/B$ as an alternative option.\footnote{\col{In fact, the ATLAS collaboration has already started to use the $c$-tagging technique in search for new physics, e.g., Refs.~\cite{Aad:2014nra,Aad:2015gna}, while the usefulness of the $c$-tagging technique in rare top decays was mentioned, e.g., Ref.~\cite{Azatov:2014lha}.} } To make a comparison of the idea in this paper with the data analyses involving the $c$-tagging technique, we again define relevant efficiencies with respect to the identification of charm-initiated jets; $\cb$ as $c$-mistagging efficiency of $b$ quark, $\cc$ as $c$-tagging efficiency of $c$ quark, and $\cs$ as $c$-mistagging efficiency of $s$ quark. Although there may exist some non-trivial correlation between heavy flavor tagging techniques \col{and the possibility of $b$-$c$ mixing tagger}~\cite{ATLASctag,Perez:2015aoa,Perez:2015lra}, to be more conservative we ignore the events in which any of the visible entities involve a contradictory result between $b$-tagging and $c$-tagging; for example, if a certain $b$ is not only $b$-tagged but $c$-tagged, the associated event is discarded. We basically require three $b$-tagged jets together with one $c$-tagged jet. Denoting the tagging scheme explained thus far as 3b1c, we have the expected numbers of signal and background events ($S_{\textnormal{3b1c}}$ and $B_{\textnormal{3b1c}}$, respectively) in this scheme as 
\bea
S_{\textnormal{3b1c}}&=& S\left\{ \mathcal{E}_b^3\mathcal{E}'_c+3\mathcal{E}_b^2\mathcal{E}_c\mathcal{E}'_b\right\} \, ,
 \label{eq:smsp} \\
B_{\textnormal{3b1c}}&=& B\left\{\mathcal{E}_b^2\mathcal{E}_s\mathcal{E}'_c+\mathcal{E}_b^2\mathcal{E}_c\mathcal{E}'_s+2\mathcal{E}_b\mathcal{E}_c\mathcal{E}_s\mathcal{E}'_b  \right\} \, ,
\label{eq:smbp} 
\eea
where $\mathcal{E}_i\equiv \textrm{b}_i(1-\textrm{c}_i)$ and $\mathcal{E}'_i\equiv \textrm{c}_i(1-\textrm{b}_i)$ $(i=b,c,s)$.

The associated $S/B$ is given by
\bea
\left(\frac{S}{B}\right)_{\textnormal{3b1c}}=\left(\frac{S}{B}\right) \frac{\mathcal{E}_b(\mathcal{E}_b\mathcal{E}'_c+3\mathcal{E}_c\mathcal{E}'_b)}{\mathcal{E}_b\mathcal{E}_s\mathcal{E}'_c+\mathcal{E}_b\mathcal{E}_c\mathcal{E}'_s+2\mathcal{E}_c\mathcal{E}_s\mathcal{E}'_b}\, ,
\eea
and in turn, the associated improvement can be  
\bea
I'=\frac{(S/B)_{\textnormal{3b1c}}}{(S/B)_{\textnormal{3b}}}=\frac{\bs}{\bb}
\left( \frac{\mathcal{E}_b(\mathcal{E}_b\mathcal{E}'_c+3\mathcal{E}_c\mathcal{E}'_b)}{\mathcal{E}_b\mathcal{E}_s\mathcal{E}'_c+\mathcal{E}_b\mathcal{E}_c\mathcal{E}'_s+2\mathcal{E}_c\mathcal{E}_s\mathcal{E}'_b} \right)\,I \, .
~~
\eea

The blue squares in FIG.~\ref{fig:improvements} demonstrate the $P_T$ dependence of $I'$. For simplicity, the efficiencies associated with $c$-tagging are taken from the corresponding average values of the medium operating point in Ref.~\cite{ATLASctag}, that is, $\cb=0.125$, $\cc=0.20$, and $\cs=0.010$. We see that $I$ is larger than $I'$ in the entire range of $P_T$. This behavior can be viewed in a simpler way by considering the leading contributions (i.e., the first terms) in eqs.~(\ref{eq:smsp}) and~(\ref{eq:smbp}). We then have
\bea
I'\approx \left(\frac{1-\cb}{1-\cs}\right)I\, ,
\eea
where $I$ is the approximated expression in eq.~(\ref{eq:Iapp}). Since $\cb$ is typically larger than $\cs$, $I'$ is typically smaller than $I$. From this series of calculations, we find that the simultaneous application of $b$- and $c$-tagging techniques is {\it not} beneficial with respect to the signal-over-background ratio. Furthermore, the expected number of signal events itself (after applying the flavor tagging techniques) becomes worse as is clear from the comparison between eqs.~(\ref{eq:sms}) and~(\ref{eq:smsp}). Thus we employ only the $b$-tagging technique for our data analysis.

Another issue that one may argue is the possibility that other SM backgrounds come into play while an additional bottom-tagged jet is required. For the cases at hand, an immediate example is $t\bar{t}b\bar{b}$ for which one $W$ gauge boson decays leptonically while the other is undetected. Certainly, this issue depends strongly on the signal processes of interest, i.e., it may not be an issue for other signal models or collider signatures. We closely look at the impact of $t\bar{t}b\bar{b}$ onto the relevant analysis, switching from the 3b scheme to the 4b scheme in conjunction with Monte Carlo simulations later.

\section{Collider study} 
\begin{table}[t]
\centering
\begin{tabular}{c|c||c|c|c}
\hline \hline
 & $m_{H^{\pm}}$ (GeV) & 3b$(\times10^{-3})$ & 4b$(\times10^{-3})$ & $I$  \\
 \hline 
$t\bar{t}$ & -- & 2.13 & 0.0350 & -- \\
\hline
$t\rightarrow ch$ & -- & 24.7 & 2.71 & 6.68\\
\hline
  & 80 & 20.4 & 2.08 & 6.21\\
  & 100 & 20.6 & 1.97 & 5.82 \\
$t\rightarrow bH^+$ & 120 & 20.2 & 2.04 & 6.15\\
 & 140 & 20.3 & 2.13 & 6.30\\
 & 160 & 20.2 & 2.11 & 6.36\\
\hline \hline
\end{tabular}
\caption{\label{tab:reduction} Reduction rates for signal and background processes in 3b and 4b schemes, and the associated improvements $I$. The reduction rates for $t\bar{t}$ are computed with all decay modes included, whereas those for signal processes are computed only with the semi-leptonic decay mode. }
\end{table}
Equipped with the rough estimate discussed thus far, we test the feasibility of the basic idea in the context of the two aforementioned example models with Monte Carlo simulations of the 14TeV LHC. 
The parton-level events for the signal and the background are generated by \texttt{MadGraph\_aMC@NLO}~\cite{Alwall:2014hca}. 
The output information is then streamed to \texttt{Pythia 6.4}~\cite{Sjostrand:2006za} and \texttt{Delphes3}~\cite{deFavereau:2013fsa} in order. Jet formation is conducted by the anti-$k_t$ algorithm with a jet radius parameter $R=0.5$. 
The $b$-tagging efficiencies in a detector simulator for bottom, charm, and light quarks are tuned according to the performance of the CSVM algorithm reported by the CMS collaboration~\cite{Chatrchyan:2012jua}. Furthermore, we apply cuts on the final state of both signal and background processes, closely following the selection scheme used in Ref.~\cite{CMSsemileptonictt} for the semi-leptonic channel of top quark pairs. The key selection criteria for leptons and jets are highlighted below with small modifications:
\begin{itemize}
\item for $e$ and $\mu$ leptons, $P_T>10$ GeV and $|\eta|<2.5$,
\item for jets, $P_T>30$ GeV and $|\eta|<4.9$,
\item for $b$-jets, $P_T>30$ GeV and $|\eta|<2.4$.
\end{itemize}

We then require different $b$-jet multiplicities for those selected events; for the 3b (4b) scheme, we exclusively demand 3 (4) $b$-tagged jets and 1 (0) regular jets together with an isolated lepton. Table~\ref{tab:reduction} summarizes not only reduction rates of signal and background processes in both schemes but the resulting improvements. Six benchmark scenarios are examined here. To see the potential dependence on the choice of the charged higgs mass, we vary it from 80 GeV to 160 GeV at intervals of 20 GeV. The reduction rates for the $t\bar{t}$ sample are evaluated with hadronic and leptonic channels included, whereas those for the signal samples are evaluated only with semi-leptonic events.

\begin{figure}[t]
\centering
\includegraphics[width=0.41\textwidth]{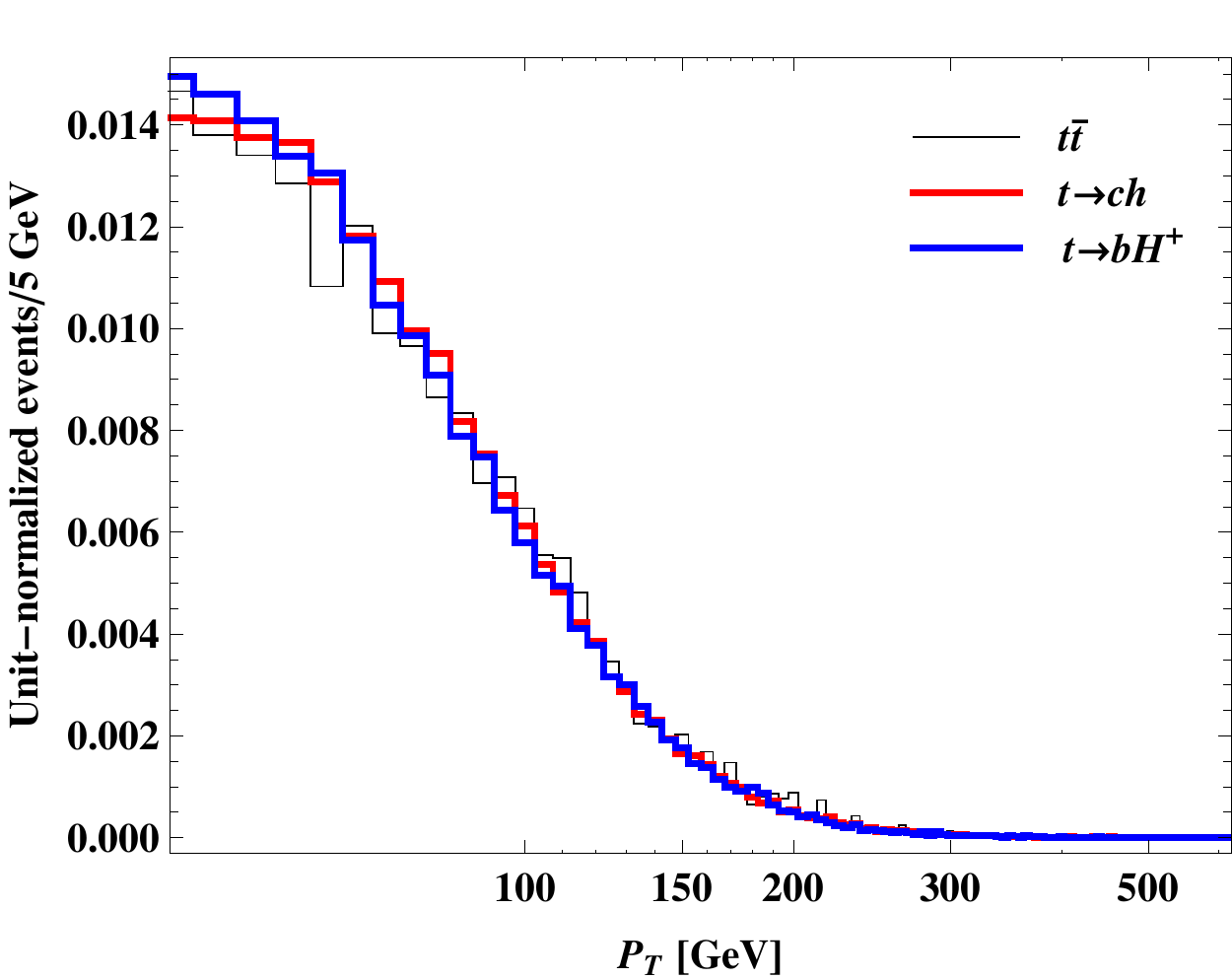}
\caption{\label{fig:pTb} The distributions of $P_T$ of bottom-tagged jets for background (black), $t\rightarrow ch$ (red), and $t\rightarrow bH^+$ (blue). For the charged higgs, a mass of 120 GeV is chosen as a representative. }
\end{figure}

We observe that the relevant signal-over-background ratio can be improved by a factor of $\mathcal{O}(6-7)$ for all scenarios; no significant dependence is shown on the choice of the mass of the charged higgs. We point out that unlike the rough estimation demonstrated in FIG.~\ref{fig:improvements}, the overall improvement is reduced by a factor of $\sim2.5$ from the maximum expected improvement. In more detail, signal efficiencies are reduced by $\sim10\%$ from 3b to 4b schemes, which does {\it not} so much differ from the mis-tagging rate for charm quarks, whereas background efficiency is degraded only by $\sim1.6\%$ that is larger than the typical mis-tagging rate for light quarks. To understand this slight mismatch, one could suspect that a large fraction of $b$-tagged jets come along with a large transverse momentum ($\gtrsim 100$ GeV) so that the overall improvement gets reduced, predicated upon the observation in FIG.~\ref{fig:improvements}. However, it turns out that only about a quarter of $b$-jets belong to this hard $P_T$ regime for both signal and background events as clear from FIG.~\ref{fig:pTb}. In fact, this is not surprising, given charged or SM-like higgs masses themselves and the mass gap between top quark and charged or SM-like higgs masses. Therefore, it does not make any substantial impact. 
We instead identify the effect from initial and final state radiations as a dominant cause of such a departure from the expectation in FIG.~\ref{fig:improvements}. More specifically, initial or final state radiated gluons, which split into a heavy flavor quark pair, play a major role because those quarks are $b$-tagged with high probability. In other words, once the contributions from $t\bar{t}b\bar{b}$ or $t\bar{t}c\bar{c}$ are switched on, the relevant efficiency drop for $t\bar{t}$ is not as large as expected, while such contributions appear as a subleading effect for signal processes. For the sake of validating this intuition, we perform another simulation of $t\bar{t}$ with initial and final state radiations completely turned off, and find that the efficiency drop between 3b and 4b schemes is restored to $\sim 0.7\%$, which is close to the typical mis-tagging rate for light quarks. 

\section{Projections and conclusions }
As mentioned at the beginning, the LHC is capable of copiously producing top quark pairs, which makes the search channels discussed here become systematics-dominated.\footnote{In case of the statistics-dominated for which the significance is evaluated by $S/\sqrt{B}$ (or, more conservatively, $S/\sqrt{S+B}$), the relevant improvement can be examined in an analogous manner to the case dealt with thus far, which is beyond the scope of this paper.}
\col{In more detail, the significance $\sigma$ is given by 
\bea
\sigma=\frac{S}{\sqrt{B+\left( \frac{\kappa}{100\%}\right)^2B^2}},
\eea
where $\kappa$ is a prefactor encoding overall systematics of backgrounds. Considering the typical $\kappa$ of $\mathcal{O}(3\%)$ for the $t\bar{t}$ channel (see, for example, Ref.~\cite{CMS:2014nta}), we find that the second term in the denominator becomes larger than the first one once $B$ is greater than $\sim 1100$. 
Note that the expected production cross section of top quark pairs at next-to-next-to-leading order including resummation of next-to-next-to-leading logarithm is 954 pb at the LHC14~\cite{Czakon:2013goa}, and the branching ratio of $t\bar{t} \rightarrow b\bar{b}c\bar{s}\ell \nu$ is $\sim 15\%$. From these two numbers, we can easily see that the number of relevant background events $B$ is much larger than $\sim 1100$ even with an integrated luminosity of 1 fb$^{-1}$.}
Therefore, the relevant significance is proportional to $S/B$ so that the improvements discussed in this letter can be directly translated into the associated signal sensitivity, that is, for a given scenario, one can probe $\sim6-7$ times smaller branching fraction into the signal process of interest than expected in the 3b scheme. Obviously, posterior analyses with kinematic variables etc. can increase $S/B$ further, which is beyond the scope of this paper. We instead leave such a research direction as future work. 

We again emphasize that the search strategy proposed here is {\it not} restricted to the benchmark scenarios employed here, but straightforwardly extended to the situations where the final state for the signal processes of interest contains charm quark-initiated jet(s) while the corresponding object(s) in backgrounds are light quark-initiated one(s). We also remark that different operating points may give rise to better improvements. Finally, we strongly encourage the ATLAS and CMS collaborations to adopt this idea as an alternative strategy in relevant new physics model searches.

\section*{Acknowledgments}
We would like to thank Richard Cavanaugh, Su-Yong Choi, Konstantin T. Matchev, Seung J. Lee, and Kohsaku Tobioka for comments and discussions. We also thank Kyoungchul Kong and Konstantin T. Matchev for a careful reading of the draft. D.K. also would like to thank CETUP* (Center for Theoretical Underground Physics and Related Areas) for its hospitality during the completion of this work. D.K. is supported by the LHC Theory Initiative postdoctoral fellowship (NSF Grant No. PHY-0969510).
M.P. is supported by the Korea Ministry of Science, ICT and Future Planning, Gyeongsangbuk-Do and 
Pohang City for Independent Junior Research Groups at the Asia Pacific Center for Theoretical Physics.
MP is also supported by World Premier International Research Center Initiative (WPI Initiative), MEXT, Japan.



\end{document}